\documentclass[twocolumn,floatfix,showkeys,preprintnumbers,amsmath,amssymb]{revtex4}

\usepackage{graphicx}
\usepackage{subfigure}
\usepackage{dcolumn}
\usepackage{bm}

\begin{document}

\title{Effect of Backing on Neutron Spectra for Low Energy Quasi-Mono-energetic p+$^7$Li Reaction.\\}
\author{H. Kumawat$^1$\footnote{author. Email address: harphool@barc.gov.in}}
\affiliation{$^1$Nuclear Physics Division, BARC, Mumbai-400085, India\\}
\date{\today}

\begin{abstract}
$\underline{\textbf{MO}}$nte-carlo $\underline{\textbf{N}}$ucleon transport $\underline{\textbf{C}}$ode (MONC) for nucleon transport is extended for below 20MeV proton transport using ENDF and EXFOR data base. It is used to simulate p+$^7$Li reaction upto 20MeV proton energies and produced neutron spectra are reported here. The simulated results are compared with calculated values from other available codes like PINO, EPEN, SimLiT and experimental data. The spectra reported here can be used to get the neutron cross-section for the quasi-mono-energetic neutron reaction and will help to subtract the low energy contribution. The neutron spectra also useful as this reaction is used for accelerator based Boron Neutron Capture Therapy. The backing materials are used to fully stop the proton beam hence contribution from the neutrons from backing materials is estimated. It is found that Tantalum is good backing material below $\sim$8 MeV and Carbon is better at higher energies. 
\end{abstract}

\maketitle
\section{\label{sec:level1}Introduction}
Measurement of neutron cross-section is an important research activity for it's application in nuclear reactors, cancer therapy, neutron dosimetry, nuclear astrophysics etc \cite{ads1, ads2,ads3,baltej,Lederer_2016,Reifarth_2014,PhysRevC.104.054608,KOEHLER1989494,Dillmann2023}. The $^7$Li(p, n)X reaction is used as quasi mono-energetic neutron source to measure cross-sections and also a possible accelerator based source of neutrons for Boron Neutron Capture Therapy (BNCT) \cite{Wang_2023,nakamura,kononov}. The threshold for the $^7$Li(p, n)$^7$Be$_g$ reaction is $\sim$1.88 MeV and cross-section rise rapidly near the threshold energy. For protons above 2.37 MeV, production of another group of neutrons starts due to inelastic state of $^7$Be at 429 keV. Neutron production threshold for three-body breakup channel $^7$Li(p, n+$^3$He)$\alpha$ is 3.7 MeV which gives a broad neutron energy distribution. Thus, $^7$Li(p, n)X reaction can be considered quasi-mono-energetic near threshold but it has a tail above 4 MeV of proton energies. It can be still used as a quasi-mono-energetic source for neutron threshold reactions where low energy tail does not contribute. The contribution of tail should be carefully subtracted for neutron capture reactions where low energy contributes more in neutron activation technique in cross-section measurement. 

Additionally, the spread in the neutron spectra occurs due to thickness of the Lithium target (up to 100 $\mu$m), used by various experimentalists and the corresponding neutron energy spread may be up to 500keV. Some experimentalists use cadmium foil to cut very low energy neutron contribution but it is important to quote the energy spectra for a particular measurement. Recently, several experiments have been conducted at BARC-FOTIA \cite{fotia1,fotia2,fotia3,BAKSHI2020162926} and BARC-TIFR \cite{PhysRevC.107.054607,MUKHERJEE201972,refId0,10149237,tifr1,Singh_2022} facilities using this reaction. MONC \cite{Kumawat:2020emd,hkvenkata13,hkroot} code is used for Monte-carlo simulations for Lithium target of thickness 4mg/cm$^2$ which is typical thickness used at many experimental facilities in Mumbai, India. The contribution of low energy tail and second peak should be considered while quoting cross-section for a single energy and in best practices it should be subtracted as mentioned in Ref. \cite{smith} or by similar methods. The neutron flux monitor reaction should be sensitive in the similar energy range as that of reaction of measurement. The calculations are also performed using code PINO \cite{pino} which includes only $^7$Li(p, n$_0$)$^7$Be$_g$ and $^7$Li(p, n$_1$)$^7$Be$^*$ reactions hence valid in low energy region ($<$ 7.0 MeV). The simulated values are also compared with available experimental data and calculated values from literature by SimLiT and EPEN \cite{epen} at 3.5 MeV.

At higher energies, the proton beam has to stop in some other material otherwise the quasi-mono-energeticity does not remain valid. Usually Tantalum and Carbon are used as the backing materials. The neutrons produced by the backing material should be estimated to get the total neutron spectrum.

The outline of this paper is as follows. In Sec. II we present brief description of MONC. Section III contains simulation results. Conclusions are
given in Sec. IV.

\section{\label{sec:level2}Brief description of MONC}
Monte Carlo program MONC incorporates Intra-nuclear Cascade, Pre-equilibrium, 
Evaporation and Fission models to simulate spallation reaction mechanism for 
thin and thick targets. Modeling details of Intra-nuclear cascade, Pre-equilibrium 
particle emission are described in detail in Ref. \cite{baras1, baras2}. Fission barrier, level density parameter and 
inverse cross sections for pre-equilibrium/evaporation/fission model are given in detail in Ref. \cite{hkumawat04, hkumawat05}.

Benchmark of spallation models for experimental values of neutron, charged particles, 
and pions double differential production cross-sections, particle multiplicities, 
spallation residues and excitation functions was organized by IAEA and is given in Ref. \cite{hkumawat10}. We have used the predecessor of this code named CASCADE.04 to calculate these quantities in the IAEA benchmark. Heat Deposition algorithm for thick spallation targets and thin films was modified and benchmarked as mentioned in Ref. \cite{hkheat08}. The code was further developed for the Neutron shielding and dosimetry applications and published \cite{hkumawat09}. Energy loss of the charge particle is calculated during the transport in the thick target.

MONC realizes the particle transport in three stages: 
1) sampling of particle (ion) mean free path in the medium taking into account the energy loss of a charged particle and a possible decay of non-stable particles ($\pi^0$, $\pi^\pm$). All $\pi^0$-mesons are considered to decay into $\gamma$-quanta at the point of their creation. The ionization losses of $\pi$ - mesons, protons and light ions are calculated by Sternheimer’s method \cite{stern} using well established Bathe formula for the average ionization loss calculations with proper density effects. Here, it is important to mention that the density effect shows reduction in ionization loss for fast charged particles due to dielectric polarization of the medium. In the lower energy region ($<$ 2.0 MeV) Lindhard’s approach \cite{lind} is used and a semi-phenomenological procedure \cite{barasion} is applied for the heavy ions. 
While doing the practical simulation one has to calculate the ionization and nuclear interaction ranges and then uses the formulation to deposit heat. 

2) Simulation of the particle interaction with a nucleus is considered along its path. 
In case of inelastic interaction the MONC code considers three stages of reaction for calculation: 
a) intranuclear cascade originally developed at Dubna: In this part of the calculation,
primary particles can be re-scattered and they may produce secondary particles several
times prior to absorption or escape from the target. Modeling of intra-nuclear cascades 
\cite{baras1, baras2} is in general rather closer to the methods used in other transport codes.
Cross-sections of the hadron-nucleus collisions are calculated based on the compilations 
of the experimental data \cite{barascrs1, barascrs2}. To calculate the nucleus-nucleus cross-sections 
we used analytical approximations with parameters defined in Ref. \cite{barascrs3}. 
b) Pre-equilibrium stage: In this part of 
the reaction, relaxation of the nuclear excitation is treated according to the 
exciton model of the pre-equilibrium decay. The relaxation is calculated by the 
method based on the Blann's model \cite{mashnik, blann}. Proton, neutron, deuterium, tritium, 
$^3$He and $^4$He are considered as emitted particles in the pre-equilibrium and in the
subsequent equilibrium stage.  c) Equilibrium stage: This part considers 
the particle evaporation/fission of the thermally equilibrated nucleus. 
\begin{figure}
\includegraphics[scale=0.52]{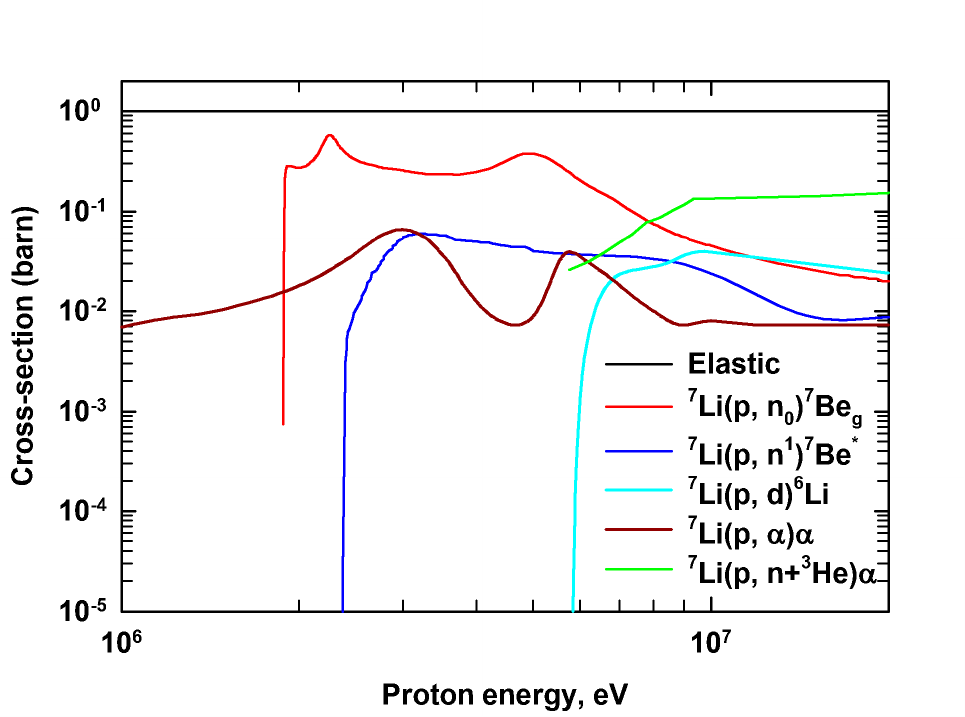}
\caption{$^7$Li(p,x)Y cross-section from ENDF VIII.0 as used in the MONC. $^7$Li(p, n+$^3$He)$\alpha$ reaction cross-section is taken from Ref. \cite{SCOTT1971405}} \label{crs}
\end{figure}

3) Low energy neutron transport code is developed recently. A package has been developed for reading pointwise cross sections for neutron in ACE (A Compact ENDF) format. ENDF data file processing and generating point data at different temperatures has also been developed \cite{endfhk}. The delayed neutrons are treated exclusively with their energy spectra for which data are available. Spontaneous and induced fission fragment yield are read from ENDF fission yield libraries. The free gas thermal treatment of the
neutron interaction for below 4eV can be used for compound and crystal material or Thermal scattering law can be used if available in ENDF file. Probability table method is used in the un-resolved energy region. Low energy ($<$ 20 MeV) proton data are used to simulate the reaction mechanism and outgoing particles energy and angular distributions. No cross-section data are given for $^7$Li(p, n+$^3$He)$\alpha$ reaction in the ENDF file so it is taken from Ref. \cite{SCOTT1971405} while energy and angular distributions are calculated using 3-body kinematics. The cross-sections used in the present simulations are given in Fig. \ref{crs} for $^7$Li nucleus. There is a thick tantalum or carbon sheet placed at the end of Lithium target to stop the proton beam. The cross-section for Tantalum and Carbon are taken from Evaluated neutron data libraries and EXFOR experimental database. The energy and angular distributions are calculated using 2, 3-body kinematics.
 \begin{figure}
 \includegraphics[scale=0.52]{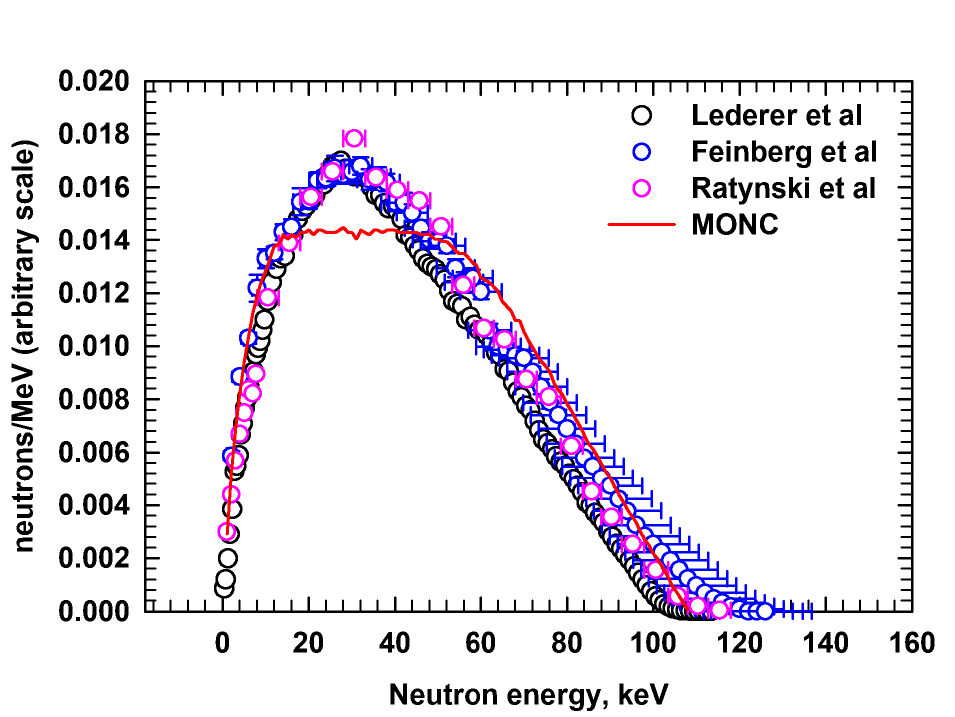}
 \caption{Proton induced neutron spectrum from $^7$Li(p,n)$^7$Be reaction. Experimental data are taken from Ref. \cite{PhysRevC.85.055809,PhysRevC.85.055810,PhysRevC.37.595} and calculations are performed using MONC.} 
 \label{sp1912}
 \end{figure}

\section{\label{sec:level3} Simulation and Results}
 
Monte-carlo simulations are carried out for 4mg/cm$^2$ thick $^7$Li target which is 92.41\% in the natural Lithium and contributes most for neutron production in the natural Lithium target. Proton energies considered are up to 21MeV (Namely 6, 10, 15 and 21 MeV). Experimental data near threshold energy at 1.912 MeV \cite {PhysRevC.85.055809,PhysRevC.85.055810,PhysRevC.37.595} are compared in Fig. \ref{sp1912}. Angular distributions are given in the ENDF library and corresponding energy is calculated using 2-body kinematics \cite{2dkin}. The calculated spectrum from MONC has overall agreement with slight underestimation at peak position of the energy spectrum.
\begin{figure}
 \includegraphics[scale=0.53]{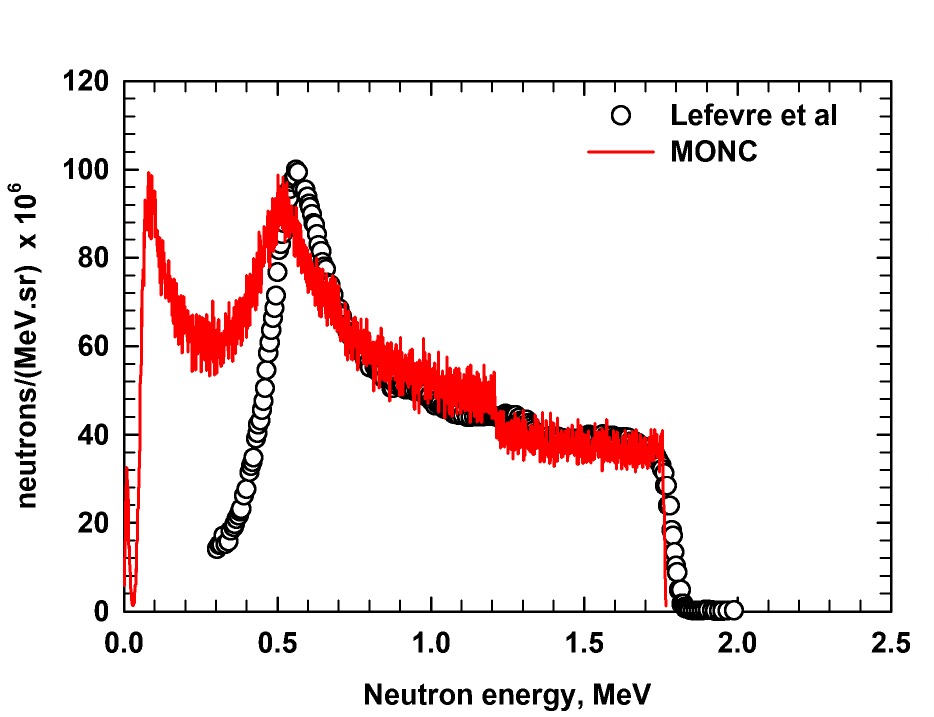}
 \caption{Proton (E$_p$=3.5 MeV) induced neutron spectrum from $^7$Li(p,n)X reaction for 0$^\circ$-5$^\circ$ of neutrons simulated using MONC. Experimental data are taken from Ref. \cite{lefevre}} 
 \label{sp352MeV}
 \end{figure}
 \begin{figure}
 \includegraphics[scale=0.52]{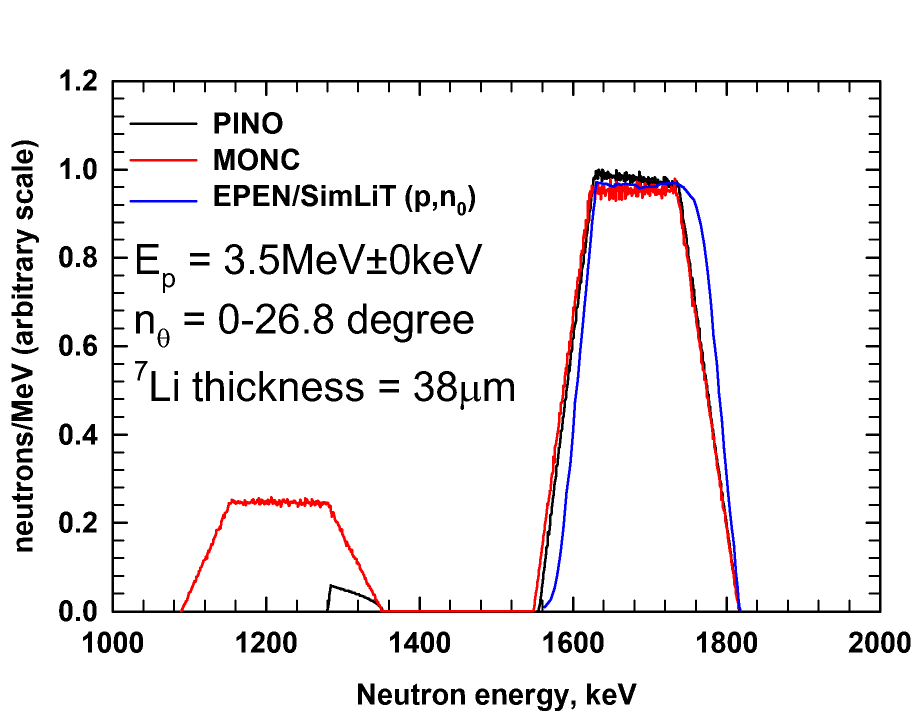}
 \caption{Proton (E$_p$=3.5 MeV) induced neutron spectrum from $^7$Li(p,n)X reaction for neutrons simulated using MONC and PINO. Values for EPEN/SimLiT are taken from Ref. \cite{epen}} 
 \label{sp35MeV}
 \end{figure}
 
\begin{figure}
 \includegraphics[scale=0.52]{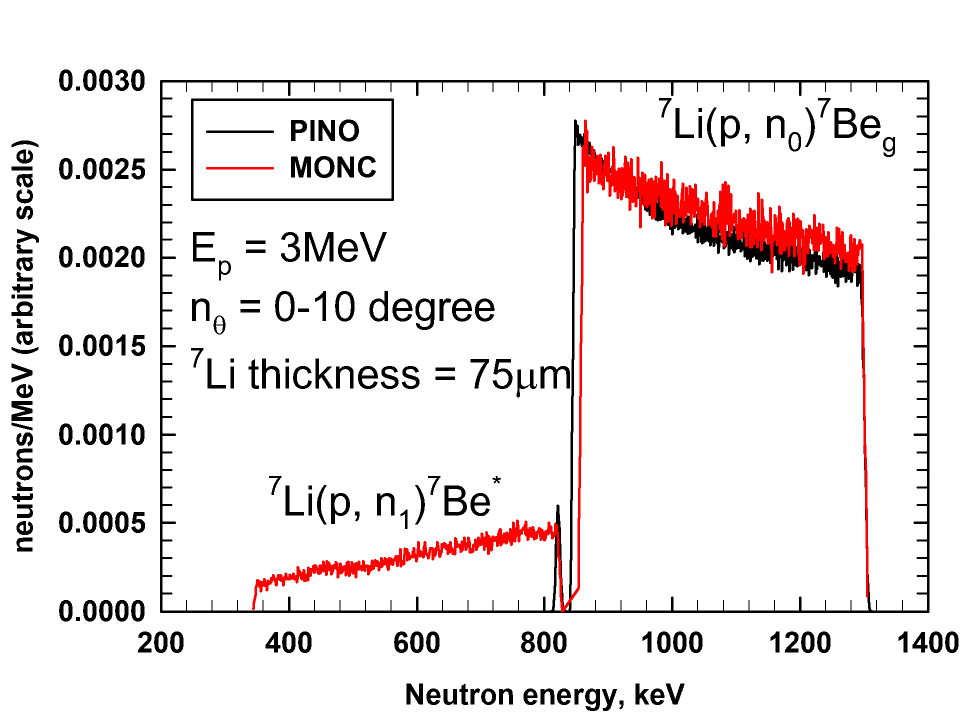}
 \caption{Proton (E$_p$=3MeV) induced neutron spectrum from $^7$Li(p,n)X reaction for 0$^\circ$-10$^\circ$ of neutrons simulated using MONC.} 
 \label{sp3MeV}
 \end{figure}

In case of thick Lithium target, experimental data at 3.45 MeV and 0$^\circ$ are taken from Ref. \cite{lefevre} and simulations are done for $\theta_n$ = 0-5$^\circ$ using MONC. A comparison is shown in Fig. \ref{sp352MeV}. There is an agreement from peak energy up to highest neutron energy. MONC overestimate the neutron spectrum below the peak neutron energy. 

Simulation was performed for 38$\mu$m thick target where EPEN and SimLiT \cite{epen} calculated values were also available in the literature. The calculations are also performed using PINO code \cite{pino} for comparison which is available on web portal. The published values from EPEN/SimLiT \cite{epen} are in good agreement with PINO and MONC calculated values for the first group of the neutron energies (see Fig. \ref{sp35MeV}) (ground state transition to $^7$Be) but the second group of neutron corresponding to the excited state transition to $^7$Be is underestimated by PINO. The data for second group of neutrons by EPEN is not given in this publication hence could not be compared.

\begin{figure*}
\includegraphics[scale=0.82]{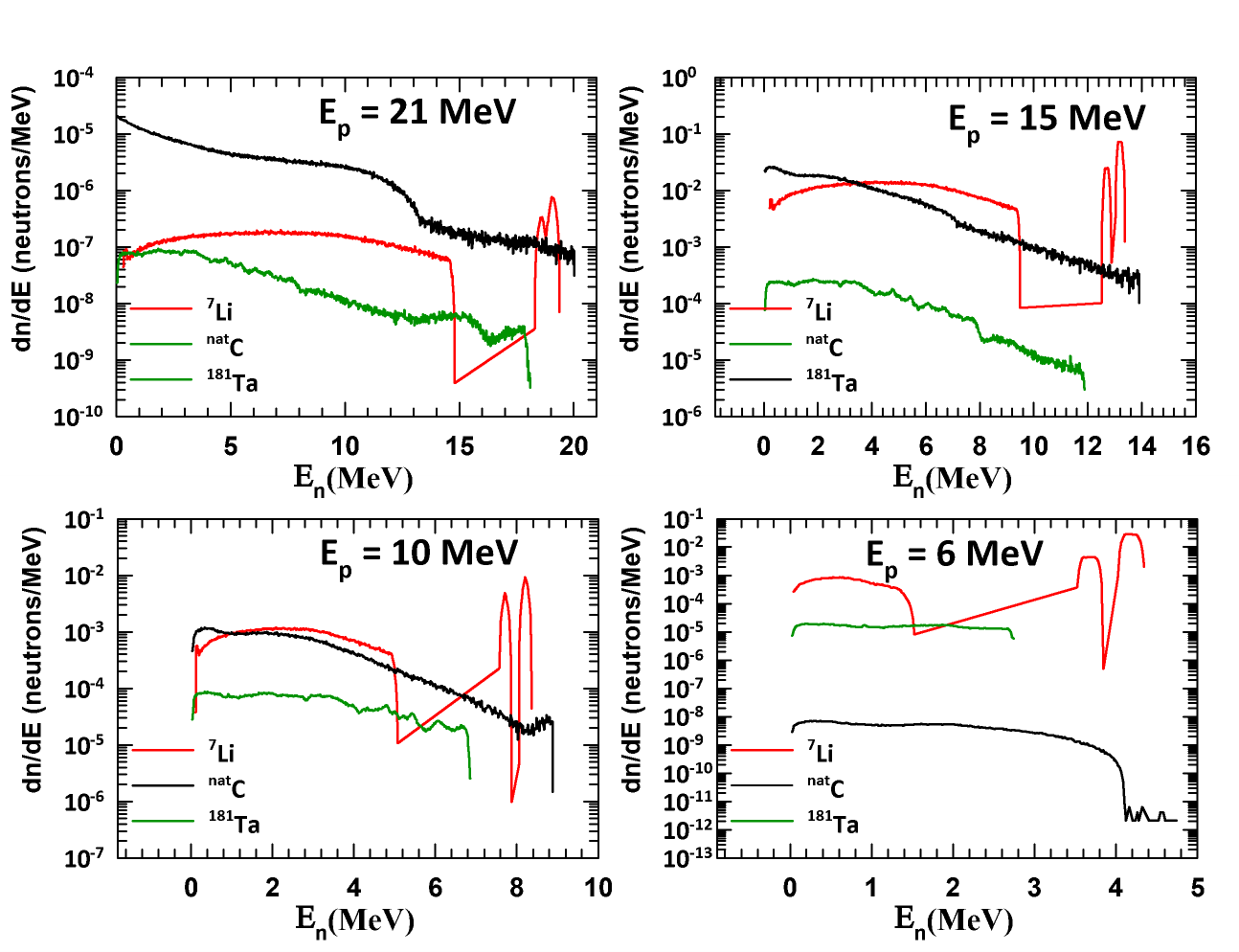}
 \caption{Proton (E$_p$=21, 15, 10 and 6 MeV) induced neutron spectrum from $^7$Li(p,n)X, $^{181}$Ta(p,n)X and $^{nat}$C(p,n)X reaction for 0$^\circ$-10$^\circ$ of neutrons simulated using MONC.} 
 \label{sp4MeV}
 \end{figure*}

Neutron spectra were simulated at 6, 10, 15 and 21 MeV proton energies. The Lithium target thickness was taken as 4mg/cm$^2$. Tantalum and Natural Carbon were taken as  backing material to stop the protons. The spectra for neutron angle $\theta_n$ 0-10$^\circ$ are given in Fig. \ref{sp4MeV} where usually samples are kept to measure neutron cross-sections. The Lithium spectra show a third group of neutrons from 6 to 20 MeV proton energies which is coming from three body breakup reaction as mentioned above. The neutrons from natural carbon target is contributed by $^{13}$C(p, n) reaction which has 1.1\% abundance. The barrier for this reaction is $\sim$1.15 MeV and threshold is 3.24 MeV. The neutron production for Tantalum is through $^{181}$Ta(p, n), $^{181}$Ta(p, 2n) and $^{181}$Ta(p, 3n) reactions with thresholds of 0.99 MeV, 7.70 MeV and 16.16 MeV, respectively. It is clear from the spectra that large contribution of neutrons is produced from Tantalum at higher energies. Carbon target gives less contribution by factor of two orders of magnitudes compared to that profuced from $^7$Li target.

\section{\label{sec:level8} Conclusion}
The Monte Carlo code MONC has been developed for proton induced reactions at low energies. The $^7$Li(p, n)X reaction which is widely used for neutron cross-section measurement and a potential candidate for Boron Neutron Capture Therapy, is investigated. The simulated neutron spectra are compared with the experimental data and calculated values from PINO and EPEN/SimLiT codes
and the results are in good agreement for the first group $^7$Li(p,n)$^7$Be$_g$ at least. There is a disagreement for the second group of neutrons through $^7$Li(p,n)$^7$Be$^*$ reaction. PINO shows very small contribution from second group of neutrons coming from excited $^7$Be state at 429 keV. Ratio of these two groups are measured at different angle \cite {PhysRev.121.871} which shows a relative contribution of 10-15\% around 0$^\circ$-10$^\circ$ neutron emission angles and the ratio from MONC are of similar magnitudes for these group of neutrons. The neutron spectra at forward angles are given here because the measurements are done in the forward direction using neutron activation technique.
The simulated neutron spectra are useful for the experimental measurements using neutron activation analysis although we use exact geometry of the arrangement of samples, sample holders, monitor foils etc. \cite{tawade} for our measurements and Monte carlo code is best suited for that compared to deterministic codes. The contribution to the neutron spectra by backing material is significant by Tantalum at higher energies and negligible below $\sim$8 MeV. The contribution from Natural carbon as backing material is less by two orders of magnitudes. The spectra from Carbon does not interfere with the major quasi-energetic peak while tantalum gives neutrons beyond this peak. Hence, Natural Carbon is better as backing material compared to Tantalum at higher energies.

  \end{document}